\documentclass[aps,prb,preprint,superscriptaddress]{revtex4-1}

\usepackage{eurosym}
\usepackage{amsfonts}
\usepackage{amssymb}
\usepackage{amsmath}
\usepackage{graphicx}
\begin{document}

\title{Nature of charge density waves and superconductivity in 1\emph{T}-TaSe$_{2-x}$Te$_x$}

\author{Y. Liu}
\email[The authors contributed equally to this work.]{}
\affiliation{Key Laboratory of Materials Physics, Institute of Solid State Physics, Chinese Academy of Sciences, Hefei 230031, People's Republic of China}

\author{D. F. Shao}
\email[The authors contributed equally to this work.]{}
\affiliation{Key Laboratory of Materials Physics, Institute of Solid State Physics, Chinese Academy of Sciences, Hefei 230031, People's Republic of China}

\author{L. J. Li}
\affiliation{Key Laboratory of Materials Physics, Institute of Solid State Physics, Chinese Academy of Sciences, Hefei 230031, People's Republic of China}

\author{W. J. Lu}
\email[Corresponding author: ]{wjlu@issp.ac.cn}
\affiliation{Key Laboratory of Materials Physics, Institute of Solid State Physics, Chinese Academy of Sciences, Hefei 230031, People's Republic of China}

\author{X. D. Zhu}
\affiliation{High Magnetic Field Laboratory, Chinese Academy of Sciences, Hefei 230031, People's Republic of China}

\author{P. Tong}
\affiliation{Key Laboratory of Materials Physics, Institute of Solid State Physics, Chinese Academy of Sciences, Hefei 230031, People's Republic of China}

\author{R. C. Xiao}
\affiliation{Key Laboratory of Materials Physics, Institute of Solid State Physics, Chinese Academy of Sciences, Hefei 230031, People's Republic of China}

\author{L. S. Ling}
\affiliation{High Magnetic Field Laboratory, Chinese Academy of Sciences, Hefei 230031, People's Republic of China}

\author{C. Y. Xi}
\affiliation{High Magnetic Field Laboratory, Chinese Academy of Sciences, Hefei 230031, People's Republic of China}

\author{L. Pi}
\affiliation{High Magnetic Field Laboratory, Chinese Academy of Sciences, Hefei 230031, People's Republic of China}

\author{H. F. Tian}
\affiliation{Beijing National Laboratory for Condensed Matter Physics, Institute of Physics, Chinese Academy of Sciences, Beijing 100190, People's Republic of China}

\author{H. X. Yang}
\email[Corresponding author: ]{hxyang@iphy.ac.cn}
\affiliation{Beijing National Laboratory for Condensed Matter Physics, Institute of Physics, Chinese Academy of Sciences, Beijing 100190, People's Republic of China}

\author{J. Q. Li}
\affiliation{Beijing National Laboratory for Condensed Matter Physics, Institute of Physics, Chinese Academy of Sciences, Beijing 100190, People's Republic of China}
\affiliation{Collaborative Innovation Center of Quantum Matter, Beijing, 100190, People's Republic of China}

\author{W. H. Song}
\affiliation{Key Laboratory of Materials Physics, Institute of Solid State Physics, Chinese Academy of Sciences, Hefei 230031, People's Republic of China}

\author{X. B. Zhu}
\affiliation{Key Laboratory of Materials Physics, Institute of Solid State Physics, Chinese Academy of Sciences, Hefei 230031, People's Republic of China}

\author{Y. P. Sun}
\email[Corresponding author: ]{ypsun@issp.ac.cn}
\affiliation{High Magnetic Field Laboratory, Chinese Academy of Sciences, Hefei 230031, People's Republic of China}
\affiliation{Key Laboratory of Materials Physics, Institute of Solid State Physics, Chinese Academy of Sciences, Hefei 230031, People's Republic of China}
\affiliation{Collaborative Innovation Centre of Advanced Microstructures, Nanjing University, Nanjing 210093, People's Republic of China}

\textbf{\begin{abstract}
Transition-metal dichalcogenides (TMDs) $MX_2$ ($M$ = Ti, Nb, Ta; $X$ = S, Se, Te) exhibit a rich set of charge density wave (CDW) orders, which usually coexist and/or compete with superconductivity. The mechanisms of CDWs and superconductivity in TMDs are still under debate. Here we perform an investigation on a typical TMD system, 1\emph{T}-TaSe$_{2-x}$Te$_x$ ($0 \leq x \leq 2$). Doping-induced disordered distribution of Se/Te suppresses CDWs in 1\emph{T}-TaSe$_2$. A domelike superconducting phase with the maximum $T_\textrm{c}^{\textrm{onset}}$ of 2.5 K was observed near CDWs. The superconducting volume is very small inside the CDW phase and becomes very large instantly when the CDW phase is fully suppressed. The observations can be understood based on the strong \emph{\textbf{q}}-dependent electron-phonon coupling-induced periodic-lattice-distortion (PLD) mechanism of CDWs. The volume variation of superconductivity implies the emergence of domain walls in the suppressing process of CDWs. Our concluded scenario makes a fundamental understanding about CDWs and related superconductivity in TMDs.
\end{abstract}}

\maketitle

\lq\lq Unconventional superconductivity\rq\rq refers superconductivity cannot be explained by the conventional electron-phonon coupling mechanism. Usually, unconventional superconductivity appears near the boundary of an ordered phase with broken translational or spin rotation symmetry \cite{Heavy-fermion-1, Cu-superconductor, Fe-superconductor}, so that it is thought to be tightly related to a (purely electronic) quantum critical point (QCP) \cite{QCP-1, QCP-2}. Transition metal dichalcogenides (TMDs) MX$_2$, where M = Ti, Nb, Ta, \emph{etc.}, and X = S, Se, Te, exhibit a rich set of Peierls-like charge density wave (CDW) orders \cite{Wilson-CDW-review}. Many typical TMDs show the coexistence and/or competition between conventional superconductivity and CDW \cite{Zhang-TaS2,Sipos-1T-TaS2-pressure, Morosan-CuxTiSe2, Morosan-PdxTiSe2, Wagner-CuxTaS2, LLJ-EPL, Ang-PRL, LY-APL, Ang-PRB, LY-JAP}. The resulted phase diagrams are very similar to those of unconventional superconductors, indicating that such superconductivity might be potentially due to a new kind of QCP unrelated to magnetic degrees of freedom \cite{TiSe2-pressure, TiSe2-QCP-1, TiSe2-QCP-2}. However, QCP is found to be far away from superconductivity recently in 1\emph{T}-TiSe$_2$ under pressure \cite{TiSe2-domain-wall}.

A better understanding of the relation needs to figure out the origin of CDW, which is a rather old but still long-standing issue in condensed matter physics \cite{Peierls-book-1955, Peierls-book-1991, Frohlich-1954}. The CDW and accompanied period-lattice-distortion (PLD) are usually explained by Peierls picture \cite{Peierls-book-1955, Peierls-book-1991, Frohlich-1954, Johannes-nesting}: Fermi surface nesting, a pure electronic effect, drives the charge redistribution regardless of whether or not PLD subsequently happens. There is an opposite mechanism that the charge redistribution is driven by strong \emph{\textbf{q}}-dependent electron-phonon coupling induced PLD, while Fermi surface nesting only plays a minor role \cite{Chan-1973, Johannes-nesting}.

The typical system 1\emph{T}-TaX$_2$ (X = S, Se, Te) (Figs.~\ref{Fig_1} (a)-(c)) is a good platform to investigate CDW and superconductivity. Some reports suggested Fermi surface nesting leads to CDW in the system \cite{Wilson-CDW-review, TaS2-xray, Myron-1975, Myron-1977, Battaglia-NbTe2, Sharma-TaX2}, while some other investigations supported PLD mechanism \cite{AmyLiu-1T-TaS2, AmyLiu-1T-TaSe2}. Making solid solution of different parent materials of the system and observing the variation of CDW vector might help to figure out the universal CDW mechanism. Moreover, the correlation effect opens Mott gap of 5\emph{d}-band of Ta in commensurate (C) CDW in 1\emph{T}-TaX$_2$ (X = S, Se) \cite{Ang-PRB}. Suppression of CCDW in 1\emph{T}-TaS$_2$ leads to a unique nearly commensurate (NC) CDW ground state, which is composed of metallic incommensurate (IC) network and Mott insulating CCDW domains \cite{Wilson-CDW-review}. Superconductivity emerges only in the percolated metallic interdomain area \cite{Sipos-1T-TaS2-pressure, LY-APL, LLJ-EPL, Ang-PRL}, which is clearly not related to the QCP. On the other hand, at low temperatures 1\emph{T}-TaSe$_2$ exhibits CCDW the same as 1\emph{T}-TaS$_2$ does \cite{Wilson-CDW-review}. However, no NCCDW phase has been found in 1\emph{T}-TaSe$_2$. Introducing superconductivity into 1\emph{T}-TaSe$_2$ and comparing it with the superconductivity in 1\emph{T}-TaS$_2$ might lead to a better understanding of the universal relation of CDW and superconductivity in TMDs.

Previously, we found that the isovalent substitution in 1\emph{T}-TaS$_{2-x}$Se$_x$ ($0 \leq x \leq 2$) system suppresses CCDW and NCCDW accompanied with superconductivity emergence in the middle doping area \cite{LY-APL, Ang-PRB, Ang-NC}. An ordered stacking of S/Ta/Se sandwiches is observed in 1\emph{T}-TaSSe \cite{Ang-NC}. In the present work, we prepared a series of 1\emph{T}-TaSe$_{2-x}$Te$_x$ ($0 \leq x \leq 2$) single crystals and obtained a phase diagram through the transport measurements. Different from the case in 1\emph{T}-TaS$_{2-x}$Se$_x$, we found that the doping induces Se/Te disorder in the system and suppresses CDW when $0.5 < x < 1.5$. A dome-like superconductivity with maximum $T_\textrm{c}^{\textrm{onset}}$ of 2.5 K was observed near CDW. The superconducting volume is very small inside the CDW phase and becomes very large instantly when CDW is fully suppressed. Our observations can be clearly understood based on PLD mechanism. The volume variation of superconductivity implies the emergence of domain walls when CDW is suppressed.

\section*{Superconducting dome near CDW}

\begin{figure}
\includegraphics[width=0.95\textwidth]{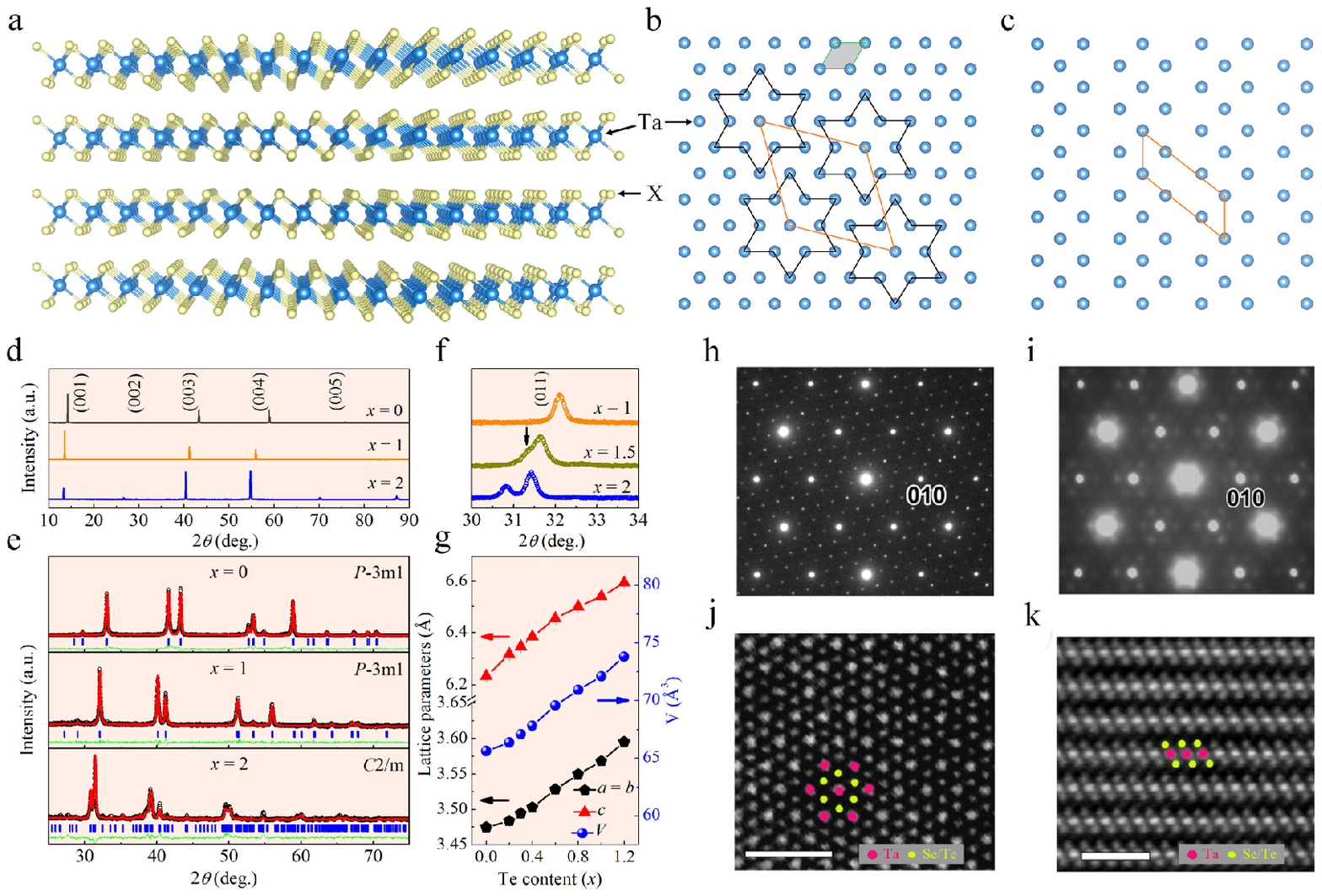}
\caption{\textbf{Structural information of 1\emph{T}-TaSe$_{2-x}$Te$_x$.} (a) Crystal structure of ideal 1\emph{T}-TaX$_2$ (X = S, Se, Te). (b) Top view of Ta plane in 1\emph{T}-TaX$_2$ (X = S, Se). At low temperatures, Ta atoms displace to make \lq\lq David star\rq\rq clusters, leading to 13.9$^\circ$ rotated $\sqrt{13}\times\sqrt{13}$ CCDW with $\emph{\textbf{q}}_{\textrm{CCDW}}=\frac{3}{13}\emph{\textbf{a}}^*+\frac{1}{13}\emph{\textbf{b}}^*$ \cite{Wilson-CDW-review}. (c) Top view of Ta plane in TaTe$_2$. The monoclinic distorted-1\emph{T} structure of TaTe$_2$ can be seen as a $3 \times 1$ single-\emph{\textbf{q}} CDW-type distorted structure of hypothetical 1\emph{T}-TaTe$_2$\cite{Wilson-CDW-review,Battaglia-NbTe2,Sharma-TaX2}. (d) Single-crystal XRD patterns of 1\emph{T}-TaSe$_{2-x}$Te$_x$ for \emph{x} = 0, 1, and 2, respectively. (e) Powder XRD patterns with Rietveld refinements of 1\emph{T}-TaSe$_{2-x}$Te$_x$ for \emph{x} = 0, 1, and 2, respectively. (f) The enlargement of the (011) peaks of the powder XRD patterns of 1\emph{T}-TaSe$_{2-x}$Te$_x$ for \emph{x} = 1, 1.5, and 2. (g) Evolution of lattice parameters (\emph{a}, \emph{c}) and cell volume (\emph{V}) of 1\emph{T}-TaSe$_{2-x}$Te$_x$. Electron diffraction patterns of (h) 1\emph{T}-TaSe$_2$, and (i) 1\emph{T}-TaSeTe, taken along the [001] zone axis direction. HAADF STEM images of 1\emph{T}-TaSeTe viewed from (j) [001] and (k) [100] zone axis direction. Scale bar, 1 nm.}
\label{Fig_1}
\end{figure}

\begin{figure}
\includegraphics[width=0.95\columnwidth]{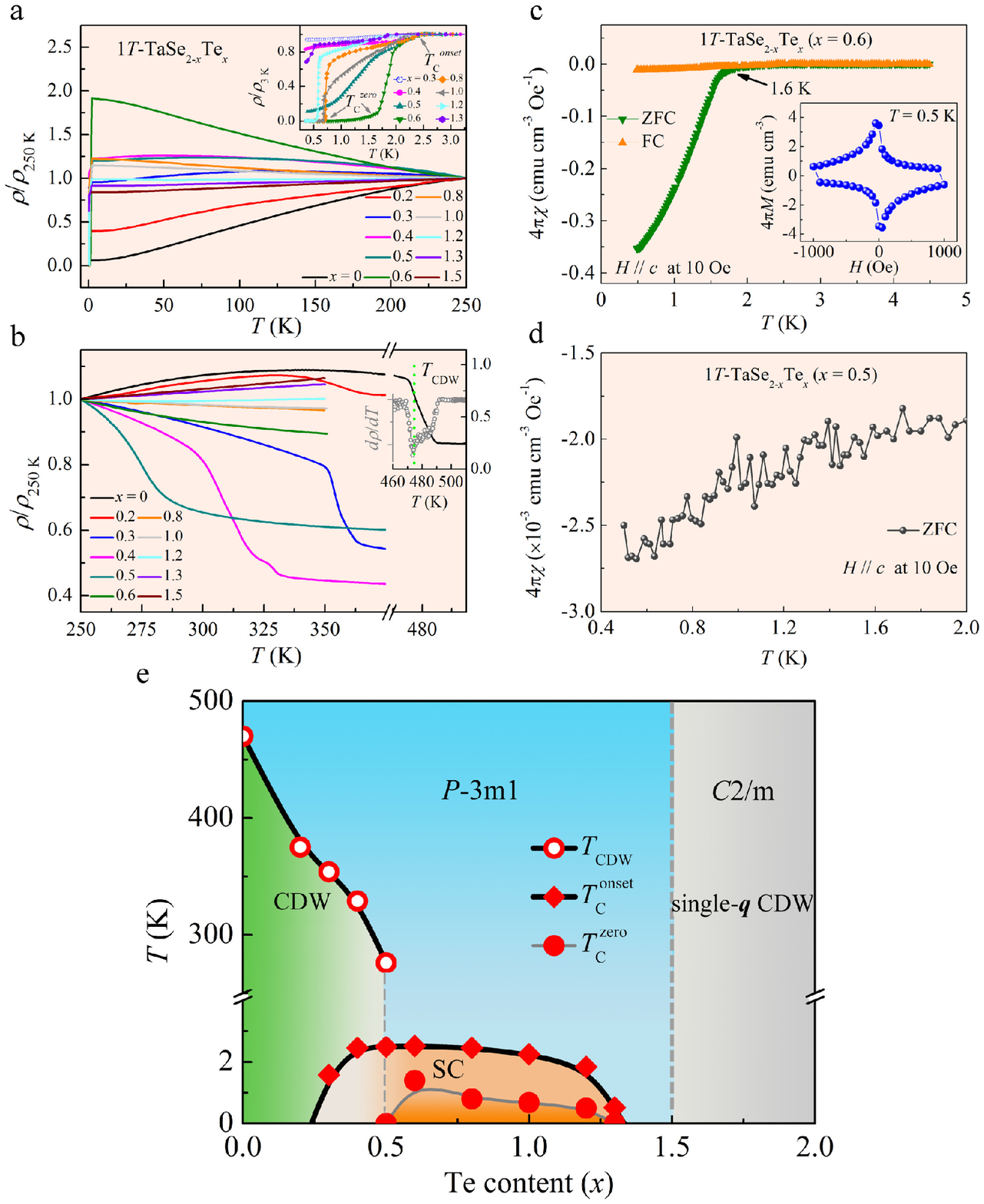}
\caption{\textbf{ The phase diagram of 1\emph{T}-TaSe$_{2-x}$Te$_x$.} Temperature dependence of in-plane resistivity ratio ($\rho/\rho_{250K}$) of 1\emph{T}-TaSe$_{2-x}$Te$_x$ below \emph{T} = 250 K (a) and above \emph{T} = 250 K (b). Insets: the enlargement of superconducting transitions at low temperatures and the CCDW transition of 1\emph{T}-TaSe$_2$ at high temperatures. Temperature dependence of magnetic susceptibility ($4\pi\chi$) for $x = 0.6$ (c) and $x = 0.5$ (d). Inset: the magnetization hysteresis loop obtained at \emph{T} = 0.5 K with magnetic field \emph{H} paralleling \emph{c}-axis. (e) Electronic phase diagram of 1\emph{T}-TaSe$_{2-x}$Te$_x$ as a function of temperature and Te content.}
\label{Fig_2}
\end{figure}

The X-ray diffraction (XRD) patterns of 1\emph{T}-TaSe$_{2-x}$Te$_x$ (\emph{x} = 0, 1, and 2) single crystals are shown in Fig.~\ref{Fig_1} (d), in which only (00l) reflections were observed, suggesting the \emph{c}-axis is perpendicular to the surface of single crystal. With increasing \emph{x}, the diffraction peaks distinctly shift to lower angles, reflecting the crystal expansion induced by Te doping. Figure~\ref{Fig_1} (e) shows the powder XRD patterns and the structural refinement results of Rietveld analysis for the selected samples with \emph{x} = 0, 1, and 2. Figure~\ref{Fig_1} (f) shows the enlargement of the (011) peak for \emph{x} = 1, 1.5, and 2. The undistorted CdI$_2$-type 1\emph{T} structure for \emph{x} = 1 leads to a single (011) peak, while there are double peaks resulting from the monoclinic distorted-1\emph{T} structure for \emph{x} = 2. The (011) peak starts to split when \emph{x} = 1.5, indicating the emergence of distorted-1\emph{T} structure, as shown in Fig.~\ref{Fig_1} (c). The evolution of lattice parameters (\emph{a}/\emph{b}, \emph{c}) and unit cell volume (\emph{V}) of 1\emph{T}-TaSe$_{2-x}$Te$_x$ are depicted in Fig.~\ref{Fig_1} (g). Indeed, the values of \emph{a}, \emph{c}, and \emph{V} monotonously increase with \emph{x}, in accordance with the larger ion radius of Te than that of Se.

Figures~\ref{Fig_1} (h) and (i) demonstrate two typical diffraction patterns taken along the [001] zone-axis direction of 1\emph{T}-TaSe$_2$ and 1\emph{T}-TaSeTe, respectively. The superstructure reflections corresponding to the commensurate phase with $\emph{\textbf{q}}_{\textrm{CCDW}}=\frac{3}{13}\emph{\textbf{a}}^*+\frac{1}{13}\emph{\textbf{b}}^*$ can be clearly observed in Fig.~\ref{Fig_1} (h), which is evidently different from the diffuse superlattice spots with incommensurated in-plane \emph{\textbf{q}} vector in 1\emph{T}-TaSeTe (Fig.~\ref{Fig_1} (i)). Similar result was previously reported by Luo \emph{et al.} \cite{Cava-PNAS} and a possible chemical origin, short-range Se/Te ordering was proposed. However, the detailed analysis based on the high-angle annular dark-field (HAADF) scanning transmission electron microscopy (STEM) images seen from both [001] (Fig.~\ref{Fig_1} (j)) and [100] (Fig.~\ref{Fig_1} (k)) zone axis directions suggest the disordered Se/Te distribution in the 1\emph{T}-TaSeTe, which is significantly different from the ordered S/Se distribution in 1\emph{T}-TaSSe\cite{Ang-NC}.

Figures~\ref{Fig_2} (a) and (b) show the temperature dependence of in-plane resistivity ratio ($\rho/\rho_{\textrm{250K}}$) of 1\emph{T}-TaSe$_{2-x}$Te$_x$ single crystals. The Te doping largely decreases the residual resistivity ratio (RRR = $\rho_{\textrm{300K}}/\rho_{\textrm{3K}}$). For 1\emph{T}-TaSe$_2$, RRR = 17.4. For 1\emph{T}-TaSeTe, RRR = 0.85, reflecting the substantial doping induced disorder, which is corresponding to our HAADF-STEM observation. The signature of superconductivity emerges as $x \geq 0.3$, and finally disappears for $x \geq 1.3$, while the zero resistances are observed when $0.5 < x < 1.3$ (Fig.~\ref{Fig_2} (a)). The maximum of superconducting onset temperature ($T_\textrm{c}^{\textrm{onset}}$) of 2.5 K and the maximum of zero resistance temperature ($T_\textrm{c}^{\textrm{zero}}$) of 1.6 K are found when CDW is completely suppressed (\emph{x} = 0.6). Figures~\ref{Fig_2} (c) and (d) show the magnetic properties of 1\emph{T}-TaSe$_{1.4}$Te$_{0.6}$ (optimal sample) and 1\emph{T}-TaSe$_{1.5}$Te$_{0.5}$ at \emph{H} = 10 Oe with the magnetic field paralleling to the \emph{c}-axis, respectively. The results indicate very small superconducting volume in 1\emph{T}-TaSe$_{1.5}$Te$_{0.5}$ (in which CDW still exists) and large superconducting volume in 1\emph{T}-TaSe$_{1.4}$Te$_{0.6}$ (in which CDW is just completely suppressed). The inset of Fig.~\ref{Fig_2} (c) shows the magnetization hysteresis loop \emph{M}(\emph{H}) obtained at \emph{T} = 0.5 K, which shows that 1\emph{T}-TaSe$_{1.4}$Te$_{0.6}$ is a typical type-II superconductor.

Figure~\ref{Fig_2} (e) summarizes the overall phase diagram as a function of temperature and doping level in 1\emph{T}-TaSe$_{2-x}$Te$_x$. The dome-like superconducting phase is near the CDW phase, which is similar to that in 1\emph{T}-TaS$_{2-x}$Se$_x$\cite{LY-APL}. In the phase diagram, the CDW is gradually suppressed by Te doping and disappears as $x > 0.5$, which is quite different from the situation in 1\emph{T}-TaS$_{2-x}$Se$_x$ \cite{LY-APL}. With higher Te content $x > 1.5$, the crystal structure gradually distorts to a monoclinic one with the \emph{C}2/m space group, which could also be considered as a single-\emph{\textbf{q}} CDW-type distortion \cite{Wilson-CDW-review, Sharma-TaX2}.

\section*{Mechanisms of CDW and superconductivity}

\begin{figure}
\includegraphics[width=0.95\textwidth]{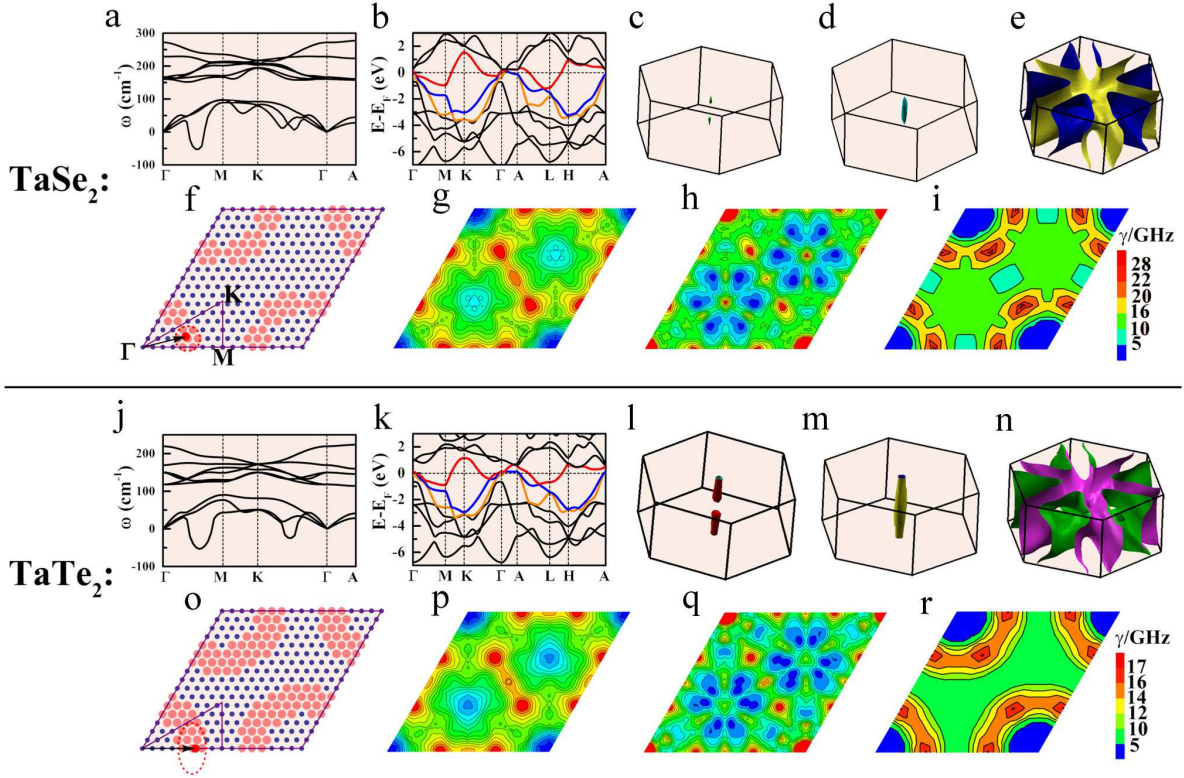}
\caption{\textbf{The first principles investigations on 1\emph{T}-TaSe$_2$ and 1\emph{T}-TaTe$_2$.} The top panel shows the phonon dispersions (a), band dispersions (b), Fermi surfaces (c)-(e), the distribution of instability in the $\emph{\textbf{q}}_z$ = 0 plane (f), the cross section of real (g) and imaginary (h) parts of the generated electron susceptibility with $\emph{\textbf{q}}_z$ = 0, and the contour map of the phonon linewidth $\gamma$ of the lowest phonon modes in the $\emph{\textbf{q}}_z$ = 0 of 1\emph{T}-TaSe$_2$ (i). The bottom panel shows those of 1\emph{T}-TaTe$_2$ ((j)-(r)). The bands crossing $E_F$ are colored in (b) and (k). In (f) and (o), the pink solid circles denote the \emph{\textbf{q}}-points at which the frequency of the lowest mode is imaginary. The red solid circles and the black arrows in (f) and (o) denote the reported $\emph{\textbf{q}}_{\textrm{CCDW}}=\frac{3}{13}\emph{\textbf{a}}^*\textrm{+}\frac{1}{13}\emph{\textbf{b}}^*$ and $\emph{\textbf{q}}\approx\frac{1}{3}\emph{\textbf{a}}^*$ for 1\emph{T}-TaSe$_2$ and 1\emph{T}-TaTe$_2$, respectively. The high-symmetry K-points are shown in (f).}
\label{Fig_3}
\end{figure}

\begin{table}
\caption{\label{Structure_LDA}Structural parameters fully optimized by LDA for 1\emph{T}-TaSe$_2$, 1\emph{T}-TaSeTe(O),and 1\emph{T}-TaTe$_2$.}
\begin{ruledtabular}
\begin{tabular}{cccc}
&\emph{a}({\AA}) &\emph{c}({\AA}) &$z_{\textrm{X}}$ \\
\hline TaSe$_2$ & 3.406 & 6.086 & $z_{\textrm{Se}}=\pm0.271$ \\
TaTe$_2$ & 3.622 & 6.572 & $z_{\textrm{Te}}=\pm0.274$ \\
TaSeTe(O) & 3.507 & 6.337 & $z_{\textrm{Se}}=0.249$,$z_{\textrm{Te}}=-0.296$ \\
\end{tabular}
\end{ruledtabular}
\end{table}

To explain our observation in 1\emph{T}-TaSe$_{2-x}$Te$_x$ system, we performed the first principle calculations (Fig.~\ref{Fig_3}). We firstly calculated the two end members of 1\emph{T}-TaSe$_2$ and hypothetical 1\emph{T}-TaTe$_2$, respectively. The fully optimized structural parameters, listed in Table~\ref{Structure_LDA}, are close to those from the previous LDA calculation \cite{AmyLiu-1T-TaSe2}. The underestimation of lattice parameters is expected for LDA \cite{AmyLiu-1T-TaSe2}.

Previous researches showed the phonon calculation is an effective method to simulate the CDW instability \cite{Johannes-NbSe2, AmyLiu-2H-TaSe2, Calandra-1T-TiSe2, Battaglia-NbTe2, AmyLiu-1T-TaS2, AmyLiu-1T-TaSe2}. Figures~\ref{Fig_3} (a) and (j) show the phonon dispersions of 1\emph{T}-TaSe$_2$ and 1\emph{T}-TaTe$_2$. For 1\emph{T}-TaSe$_2$, the calculation is in good agreement with the previous calculation by Ge \emph{et al}. \cite{AmyLiu-1T-TaSe2}. Form Fig.~\ref{Fig_3} (f), one can notice that the calculated instability is just around the reported CCDW vector ($\emph{\textbf{q}}_{\textrm{CCDW}}=\frac{3}{13}\emph{\textbf{a}}^*+\frac{1}{13}\emph{\textbf{b}}^*$) \cite{TaS2-xray}. For 1\emph{T}-TaTe$_2$, the area of instability is centered near $\emph{\textbf{q}}_{\textrm{CDW}}\approx\frac{1}{3}\emph{\textbf{a}}^*$ (Fig.~\ref{Fig_3} (o)), which is corresponding to the reported ($3 \times1 $) single-\textbf{\emph{q}} CDW-type superlattice \cite{Battaglia-NbTe2}. The high coincidence of the calculated and experimentally reported instabilities strongly proves the reliability and accuracy of the phonon calculation. Moreover, different from 1\emph{T}-TaSe$_2$, 1\emph{T}-TaTe$_2$ shows a much larger area of instability, which expands to $\Gamma$K line (see Fig.~\ref{Fig_3} (j)). That might be the reason why 1\emph{T}-TaSe$_2$ only shows the small atomic displacement in CDW phase, which can be suppressed at high temperatures, while the single-\emph{\textbf{q}} CDW-type distortion in TaTe$_2$ is very stable and the ideal-1\emph{T} structure has never been observed.

Figures~\ref{Fig_3} (b) and (k) show the band structures and Fermi surfaces of 1\emph{T}-TaSe$_2$ and 1\emph{T}-TaTe$_2$, respectively. For 1\emph{T}-TaSe$_2$, early calculations show there is only one band crossing Fermi energy ($E_\textrm{F}$), which does not cross $E_\textrm{F}$ in the vicinity of $\Gamma$ point \cite{Myron-1975, Woolley-1977}. Moreover, there is a gap (about $0.1 \sim 0.2$ eV) below the band crossing $E_\textrm{F}$ \cite{Myron-1975, Woolley-1977}. However, the recent angle-resolved photoemission (ARPES) experiment clearly shows a hybridization of bands at $\Gamma$ close to $E_\textrm{F}$, where small hole-type pockets are observed \cite{Bovet-ARPES}. Obviously, our LDA calculations accurately simulated the band structure of 1\emph{T}-TaSe$_2$ (Fig.~\ref{Fig_3} (b)). Three bands cross $E_\textrm{F}$: The lower two bands (colored in blue and orange in Fig.~\ref{Fig_3} (b)) form small cylindrical hole-type pockets close to $\Gamma$ (Figs.~\ref{Fig_3} (c) and (d)). The Fermi surface introduced from the higher band crossing $E_\textrm{F}$ (colored in red in Fig.~\ref{Fig_3} (b)) is shown in Fig.~\ref{Fig_3} (e). 1\emph{T}-TaTe$_2$, one can notice that the band structure and Fermi surfaces (Figs.~\ref{Fig_3} (k) - (n)) are highly similar to those of 1\emph{T}-TaSe$_2$.

The Fermi surface nesting can be reflected in generated electron susceptibility \cite{Johannes-nesting}. Figures~\ref{Fig_3} (g), (h), (p) and (q) show the cross section of the real part ($\chi^{'}$) and imaginary part ($\chi^{''}$) of the electron susceptibility with $\emph{\textbf{q}}_\textrm{z} = 0$ for 1\emph{T}-TaSe$_2$ and 1\emph{T}-TaTe$_2$. We found that all the maxima of $\chi^{'}$ and $\chi^{''}$ locate between $\Gamma$ and M points. For 1\emph{T}-TaSe$_2$, both the maxima of $\chi^{'}$ and $\chi^{''}$ locate at $\emph{\textbf{q}}\approx\frac{1}{3}\emph{\textbf{a}}^*$. Earlier calculation by Myron \emph{et al}. \cite{Myron-1977} shows a peak of $\chi^{'}$ at $\emph{\textbf{q}}\approx 0.28\emph{\textbf{a}}^*$, while recent calculation by Yu \emph{et al}.\cite{Yu-arxiv} reports a maximum of $\chi^{'}$ at $\emph{\textbf{q}}\approx 0.295\emph{\textbf{a}}^*$. Clearly, the maxima of $\chi^{'}$ and $\chi^{''}$ locate far away from $\emph{\textbf{q}}_{\textrm{CDW}}=\frac{3}{13}\emph{\textbf{a}}^*+\frac{1}{13}\emph{\textbf{b}}^*$. Therefore, Fermi surface nesting cannot account for the origin of CDW in 1\emph{T}-TaSe$_2$.

We also calculated the electron-phonon coupling in the $\emph{\textbf{q}}_\textrm{z} = 0$ plane for 1\emph{T}-TaSe$_2$ and 1\emph{T}-TaTe$_2$. Figures~\ref{Fig_3} (i) and (r) show the calculated phonon linewidth $\gamma$ of the lowest phonon modes in the $\emph{\textbf{q}}_\textrm{z} = 0$ plane. Although the calculation with $16 \times 16 \times 1$ \emph{\textbf{q}}-points is not enough to deduce the accurate vector with maximum $\gamma$, it can still qualitatively reflect the role of electron-phonon coupling. In the instability area, the $\gamma$ of the lowest mode is hundreds times larger than those of higher modes, proving the connection between electron-phonon coupling and CDW. For 1\emph{T}-TaSe$_2$, the biggest $\gamma$ ($\sim$ 23.11 GHz) is found near the place where $\chi^{''}$ shows the maximum, which is understandable since Fermi surface nesting can enhance $\gamma$ according to Eqs.(\ref{Eq_gammar}) and (\ref{Eq_chi}) (Methods). The second largest $\gamma$ ($\sim$ 22.18 GHz) is found in the place very near the reported $\emph{\textbf{q}}_{\textrm{CDW}}=\frac{3}{13}\emph{\textbf{a}}^*+\frac{1}{13}\emph{\textbf{b}}^*$. Therefore, if we neglect the enhancement of nesting, one can find the area of \emph{\textbf{q}}-points with large $\gamma$ is centered at the reported $\emph{\textbf{q}}_{\textrm{CDW}}$. For 1\emph{T}-TaTe$_2$, the large $\gamma$ area is strongly broadened and expands to $\Gamma$K, which is coinciding with the phonon instability area shown in Fig.~\ref{Fig_3}. Meanwhile, $\chi^{'}$ and $\chi^{''}$ show small value in the place between $\Gamma$ and K. Therefore, we can conclude that the \emph{\textbf{q}}-dependent electron-phonon coupling induced PLD, instead of Fermi surface nesting, is responsible for CDW in 1\emph{T}-TaSe$_{2-x}$Te$_x$ system.

\begin{figure}
\includegraphics[width=0.6\textwidth]{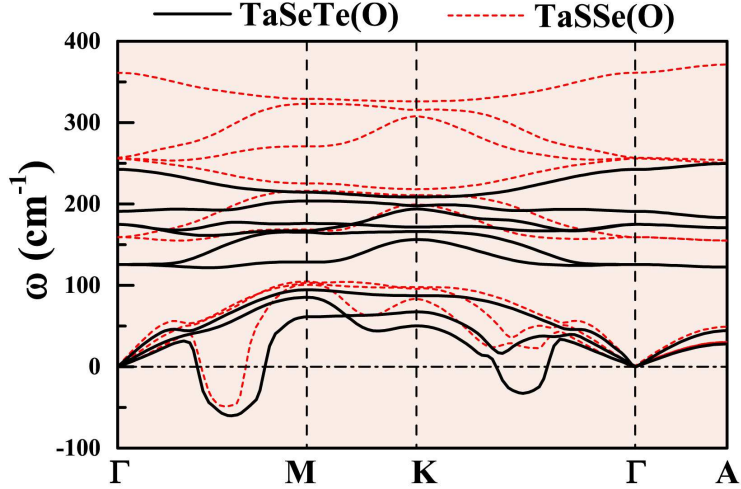}
\caption{ \textbf{Phonon dispersions of 1\emph{T}-TaSeTe(O) and 1\emph{T}-TaSSe(O).}}
\label{Fig_4}
\end{figure}

In order to explain the suppression of CDW upon Te doping, we calculated the simplest hypothetical sample 1\emph{T}-TaSeTe with an ordered stacking of Se/Ta/Te, which is represented as 1\emph{T}-TaSeTe(O). For comparison, we calculated the isostructural 1\emph{T}-TaSSe(O), in which the ordered stacking of S/Ta/Se is experimentally demonstrated \cite{Ang-NC}. Both the phonon dispersions of 1\emph{T}-TaSeTe(O) and 1\emph{T}-TaSSe(O) show the CDW instability (Fig.~\ref{Fig_4}), which means if the S/Se or Se/Te are orderly distributed, CDW could not be suppressed. Therefore, our observed suppression of CDW is due to the doping induced disorder. From Table~\ref{Structure_LDA}, one can notice that the optimized \emph{z}-coordinates of X atoms in pristine 1\emph{T}-TaX$_2$ (X = Se, Te) is about $\pm 0.27$. However, for 1\emph{T}-TaSeTe(O), the \emph{z}-coordinates of X atoms change to $z_{\textrm{Se}} = 0.249$ and $z_{\textrm{Te}} = -0.296$, which indicates the TaX$_6$ octahedra are largely distorted. When the Se and Te atoms are randomly mixed, random distortions of TaX$_6$ octahedra can be expected in reality, leading to the puckered Ta-Ta layers. This is not compatible with pure two-dimensional PLD. The above scenario can account for the fact that the disorder completely suppresses CDW in 1\emph{T}-TaSe$_{2-x}$Te$_x$ system.

We also try to understand the superconductivity near CDW based on the PLD mechanism. The electron-phonon coupling strength for each mode ($\lambda_{\textbf{\emph{q}}\nu}$) is defined as,
\begin{equation}
\lambda_{\textbf{\emph{q}}\nu}=\frac{\gamma_{\textbf{\emph{q}}\nu}}{\pi\hbar N(e_\textbf{\emph{F}})\omega^2_{\textbf{\emph{q}}\nu}}.
\label{Eq_lambda}
\end{equation}
An imaginary frequency $\omega$ of the phonon mode indicates the dynamical instability (in our cases it indicates the CDW distortion). When the CDW is suppressed, the stabilizing of 1\emph{T} structure will make the imaginary frequency $\omega$ around $\textbf{\emph{q}}_{\textrm{CDW}}$ become a small real value \cite{AmyLiu-1T-TaS2, AmyLiu-1T-TaSe2}. The large $\gamma$ and small real $\omega$ in Eq.(\ref{Eq_lambda}) can cause a large electron-phonon coupling constant, leading to the superconductivity. However, it is still hard to understand the part of superconducting phase locating inside the CDW area of the phase diagram.

\section*{Universal schematic phase diagrams}

\begin{figure}
\includegraphics[width=0.95\textwidth]{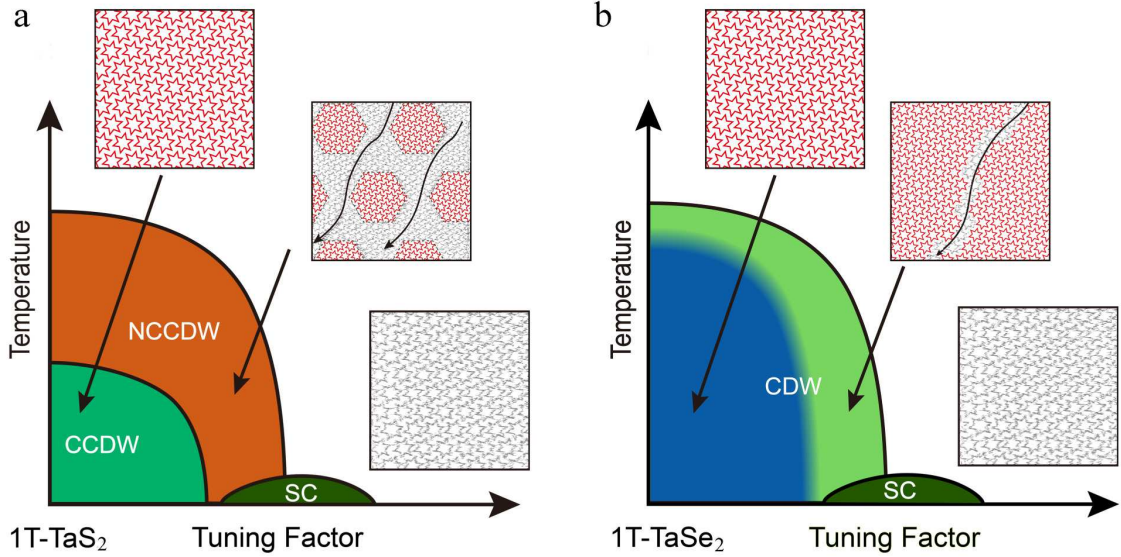}
\caption{\textbf{Universal schematic phase diagrams.} The schematic tuning phase diagrams of (a) 1\emph{T}-TaS$_2$ and (b) 1\emph{T}-TaSe$_2$.}
\label{Fig_5}
\end{figure}

1\emph{T}-TaS$_2$ and 1\emph{T}-TaSe$_2$ have the same CCDW ground state, but the CDW transitions in them are different. For 1\emph{T}-TaSe$_2$, just like most TMDs, ICCDW seems to emerge only when CCDW is fully suppressed. On the contrary, in 1\emph{T}-TaS$_2$ when CCDW is suppressed, domain walls instantly emerge and cut the previous long range CCDW into CCDW domains, leading to NCCDW state. The CCDW domains gradually shrink, while ICCDW grows in interdomain areas. When the CCDW domains disappear, the system transforms to ICCDW state. That characteristic makes 1\emph{T}-TaS$_2$ very unique among the TMDs with CDW in a long time \cite{Wilson-CDW-review}.

In CCDW states of 1\emph{T}-TaS$_2$ and 1\emph{T}-TaSe$_2$, the 5\emph{d} band of Ta opens a Mott gap, which should prohibit the superconductivity inside CCDW phase. In NCCDW state of 1\emph{T}-TaS$_2$, the CCDW domains remain Mott insulating, superconductivity can only emerge in the metallic interdomain area\cite{Sipos-1T-TaS2-pressure}. In our present 1\emph{T}-TaSe$_{2-x}$Te$_x$, considering the observed very small superconducting volume inside the CDW phase, one can expect that the domain walls emerge when CDW is suppressed, just like the case demonstrated in 1\emph{T}-TiSe$_2$ under pressure \cite{TiSe2-domain-wall}. Very recently, the coexistence of CCDW and ICCDW phases was observed during the photoinduced suppression of CDW in 1\emph{T}-TaSe$_2$\cite{YHX-PRB}, which supports our concluded scenario. Since dome-like superconductivity is usually found in TMDs with CDW, the domain walls should be universal, although they are hard to be reflected in the routine measurements.

Thus we can illustrate the universal natures of CDW and related superconductivity in some novel TMDs such as 1\emph{T}-TaS$_2$, 1\emph{T}-TaSe$_2$ and 1\emph{T}-TiSe$_2$. At $\emph{\textbf{q}}_{\textrm{CDW}}$, the strong electron-phonon coupling largely softens the phonon modes. Below $T_{\textrm{CDW}}$ the phonon energy at $\emph{\textbf{q}}_{\textrm{CDW}}$ becomes imaginary, meaning there is a new lattice structure. Since CDW originates from the strong \emph{\textbf{q}}-dependent electron-phonon coupling induced PLD instead of the Fermi surface nesting, the CDW gap does not need to be opened exactly at $E_\textrm{F}$ when the CDW transition happens. Therefore, there is no reason to have a metal-insulator transition associated with the CDW transition\cite{PNAS-CDW-class}.

We drew the schematic phase diagrams to show the tuning process of CDW in those systems (Fig.~\ref{Fig_5}). The tuning factor such as doping or pressure can harden the phonon mode at $\emph{\textbf{q}}_{\textrm{CDW}}$. When the phonon energy at $\emph{\textbf{q}}_{\textrm{CDW}}$ becomes real, CDW is suppressed. The suppression probably firstly happens in a small area, leading to the emergence of domain walls. In 1\emph{T}-TaS$_2$, the interdomain areas grow and the domains shrink upon tuning. When the domains disappear, the CDW is fully suppressed (Fig.~\ref{Fig_5} (a)). For 1\emph{T}-TaSe$_2$ and 1\emph{T}-TiSe$_2$, in CDW order the domain walls are strongly pinned. In that case, the volumes of the filament-like interdomain areas are very small, so that the routine phase diagram (Fig.~\ref{Fig_2} (e)) based on transport measurements cannot reflect the situation. A more accurate phase diagram obtained by some measurements with high resolution should follow the schematic diagram presented in Fig.~\ref{Fig_5} (b). The mechanism of different pinning of domain walls in 1\emph{T}-TaSe$_2$ and other systems needs to be further investigated.

The superconductivity near CDW is not due to QCP, but to the emergence of domain walls. In the interdomain areas, as CDW is just suppressed, the phonon frequency at $\emph{\textbf{q}}_{\textrm{CDW}}$ has a small real value, leading to a large total electron-phonon coupling constant. Once the interdomain areas percolated, superconductivity emerges (Fig.~\ref{Fig_5}). When system is far away from CDW, the phonon frequency at $\emph{\textbf{q}}_{\textrm{CDW}}$ become larger, and thus the total electron phonon coupling constant decreases. Therefore, a dome-like superconducting phase can be obtained.

In conclusion, we prepared a series of 1\emph{T}-TaSe$_{2-x}$Te$_x$ ($0 \leq x \leq 2$) single crystals and summarized an overall electronic phase diagram through the transport measurements. The CDW in 1\emph{T}-TaSe$_{2-x}$Te$_x$ ($0 \leq x \leq 2$) is gradually suppressed by Te doping induced Se/Te disorder and finally disappears when $x > 0.5$. A dome-like superconducting phase with the maximum $T_{c}^{onset}$ of 2.5 K is observed near CDW. The superconducting volume is very small inside the CDW phase and becomes very large instantly when CDW is fully suppressed. Based on the first principle calculations, we found that the origin of CDW in the system should be the strong \emph{\textbf{q}}-dependent electron-phonon coupling induced PLD instead of Fermi surface nesting. In this framework the natures of CDW and superconductivity can be well understood. The volume variation of superconductivity implies the emergence of domain walls in the suppressing process of CDW. Our concluded scenario makes a fundamental understanding about CDW and related superconductivity in TMDs.

\section*{Methods}
The single crystals were grown via the chemical vapor transport (CVT) method with iodine as a transport agent. The high-purity elements Ta (4N), Se(4N), and Te(5N) were mixed in chemical stoichiometry, and heated at 900$^\circ$C for 4 days in an evacuated quartz tube. The harvested TaSe$_{2-x}$Te$_x$ powders and iodine (density: 5 mg/cm$^3$) were then sealed in another quartz tube and heated for two weeks in a two-zone furnace, where the temperature of source and growth zones were fixed at 950$^\circ$C and 850$^\circ$C, respectively. Finally, the tubes were rapidly quenched in cold water to ensure retaining of the 1\emph{T} phase.

The X-ray diffraction (XRD) patterns were obtained on a Philips X$^\prime$pert PRO diffractometer with Cu $K_\alpha$ radiation ($\lambda$ = 1.5418 {\AA}). Structural refinements were performed by using Rietveld method with the X$^\prime$pert HighScore Plus software. Electron diffraction and HAADF-STEM experiments were performed in the JEOL ARM200F equipped with double aberration correctors and cold field emission gun operated at 200 kV. For the STEM images, the convergence angle is 28 mrad and the collection angle of the HAADF detector is between 90 and 370 mrad. Under this condition, the spatial resolution is about 0.08 nm. The electrical resistivity was measured in the \emph{ab} plane of each crystal in zero applied magnetic field by the standard four-probe ac technique using the Quantum Design Physical Property Measurement System (PPMS). The four 50 $\mu$m diameter platinum wire electrical leads were attached to the crystals$\rq$ tape-tripped fresh surface using silver epoxy. Measurement of the temperature dependence of magnetic susceptibility and the isothermal hysteresis loop were carried out in the Quantum Design Magnetic Property Measurement System (MPMS) equipped with a $^3$He cryostat.

The first principles calculations based on density functional theory (DFT) were carried out using QUANTUM ESPRESSO package \cite{QE} with ultrasoft pseudopotentials. The exchange-correlation interaction was treated with the local-density-approximation (LDA) according to Perdew and Zunger \cite{PZ}. The energy cutoff for the plane-wave basis set was 35 Ry. Brillouin zone sampling is performed on the Monkhorst-Pack (MP) mesh \cite{MP} of $32 \times 32 \times 8$. The Vanderbilt-Marzari Fermi smearing method with a smearing parameter of $\sigma=0.02$ Ry was used for the calculations of the total energy and electron charge density. Phonon dispersions were calculated using DFPT \cite{DFPT} with an $8 \times 8 \times 4$ mesh of \emph{\textbf{q}}-points. In order to investigate the distribution of CDW instability around the \textbf{\emph{q}}$_\textrm{CDW}$, $16 \times 16 \times 1$ \emph{\textbf{q}}-points were used. Denser $64 \times 64 \times 8$ \emph{\textbf{k}}-points are used for electron-phonon coupling.

The real part of the electron susceptibility is defined as
\begin{equation}
\chi^{\prime}(\emph{\textbf{q}})=\sum_\emph{\textbf{k}}\frac{f(\varepsilon_\emph{\textbf{k}})-f(\varepsilon_{\emph{\textbf{k}}\textrm{+}\emph{\textbf{q}}})}{\varepsilon_\emph{\textbf{k}}-\varepsilon_{\emph{\textbf{k}}\textrm{+}\emph{\textbf{q}}}}, \end{equation}
where $f(\varepsilon_\textbf{\emph{k}})$ is Fermi-Dirac function. The imaginary part is \cite{Johannes-nesting}
\begin{equation}
\chi^{\prime\prime}(\emph{\textbf{q}})=\sum_\emph{\textbf{k}}\delta(\varepsilon_\emph{\textbf{k}}-\varepsilon_\emph{\textbf{F}})\delta(\varepsilon_{\emph{\textbf{k}}\textrm{+}\emph{\textbf{q}}}-\varepsilon_\emph{\textbf{F}}).
\end{equation}
We used a mesh of approximately 40,000 \emph{\textbf{k}} points in the full reciprocal unit cell to calculate the energy eigenvalues derived for the electron susceptibilities.

The phonon linewidth $\gamma$ is defined by
\begin{eqnarray}
\gamma_{\textbf{\emph{q}}\nu}=2\pi\omega_{\textbf{\emph{q}}\nu}\sum_{ij}\int\frac{d^3k}{\Omega_{BZ}}|g_{\textbf{\emph{q}}\nu}(\textbf{\emph{k}},i,j)|^2\times\delta(\varepsilon_{\textbf{\emph{q}},i}-\varepsilon_\textbf{\emph{F}})\delta(\varepsilon_{\textbf{\emph{k}}+\textbf{\emph{q}},j}-\varepsilon_\textbf{\emph{F}}),
\label{Eq_gammar}
\end{eqnarray}
where the electron-phonon coefficients $g_{\textbf{\emph{q}}\nu}(\textbf{\emph{k}},i,j)$ are defined as,
\begin{equation}
g_{\textbf{\emph{q}}\nu}(\textbf{\emph{k}},i,j)=(\frac{\hbar}{2M\omega_{\textbf{\emph{q}}\nu}})^{1/2}\langle\psi_{i,\textbf{\emph{k}}}|\frac{dV_{\textrm{SCF}}}{d\hat{u}_{\textbf{\emph{q}}\nu}}\cdot\hat{\epsilon}_{\textbf{\emph{q}}\nu}|\psi_{j,\textbf{\emph{k}}+\textbf{\emph{q}}}\rangle.
\label{Eq_chi}
\end{equation}
According to this definition, $\gamma$, which reflects the electron-phonon coupling contribution, is a quantity that does not depend on real or imaginary nature of the phonon frequency.

\section*{Acknowledgements}
This work was supported by the National Key Basic Research under Contract No. 2011CBA00111, the National Nature Science Foundation of China under Contract Nos. 11404342, 11274311 and 11190022, the National Basic Research Program of China under Contract 2015CB921300, the Joint Funds of the National Natural Science Foundation of China and the Chinese Academy of Sciences¡¯ Large-scale Scientific Facility (Grand No. U1232139), Anhui Provincial Natural Science Foundation under Contract No. 1408085MA11, and the Director¡¯s Fund under Contract No.YZJJ201311 of Hefei Institutes of Physical Science, Chinese Academy of Sciences.

\section*{Author Contributions}
Y.L., L.J.L. and Y.P.S. synthesized the single crystals and carried out the XRD measurement. H.F.T., H.X.Y. and J.Q.L. performed the HAADF-STEM experiments. Y.L., X.D.Z. C.Y.X., L.P., P.T., W.H.S., L.S.L., X.B.Z. and Y.P.S performed the transport and magnetization measurements. D.F.S., R.C.X. and W.J.L. carried out the DFT calculations and explain the experimental observation. Y.L., D.F.S., W.J.L. and Y.P.S. drafted the manuscript. All authors reviewed and approved the manuscript.

\section*{Competing financial interests}
The authors declare no competing interests.

\end{document}